# Pyrolysis kinetics of Melon (*Citrullus colocynthis L.*) seed husk

Bemgba Bevan Nyakuma

*Centre for Hydrogen Energy, Institute of Future Energy,
Universiti Teknologi Malaysia, 81310 UTM Skudai, Johor Bahru, Malaysia.*

*Corresponding author Email: bbnyax1@gmail.com, bnbevan2@live.utm.my

**Abstract**

This study is aimed at investigating the thermochemical fuel characteristics and kinetic decomposition of melon seed husks (MSH) under inert (pyrolysis) conditions. The calorific value, elemental composition, proximate analyses and thermal kinetics of MSH was examined. The kinetic parameters; activation energy $E$ and frequency factor $A$ for MSH decomposition under pyrolysis conditions were determined using the *Kissinger* and isoconversional *Flynn-Wall-Ozawa* (FWO) methods. The values of $E$ for MSH ranged from 146.81 to 296 kJ/mol at degrees of conversion $\alpha$ = 0.15 to 0.60 for FWO. The decomposition of MSH process was fastest at $\alpha$ = 0.15 and slowest at $\alpha$ = 0.60 with average $E$ and $A$ values of 192.96 kJ/mol and $2.86 \times 10^{26}$ min$^{-1}$, respectively at correlation values of 0.9847. The kinetic values of MSH using the *Kissinger* method are $E$ = 161.26 kJ/mol and frequency factor, $A = 2.08 \times 10^{10}$ min$^{-1}$ with the correlation value, $R^2$ = 0.9958. The results indicate that MSH possesses important characteristics of a potential solid biofuel (SBF) for future thermochemical applications in clean energy and power generation.

**Keywords**: Biofuel, Pyrolysis, Kinetics, Melon, Seed, Husks

## Introduction

Melon (*Citrullus colocynthis L.*) is an important oil seed and perennial cash crop widely cultivated in many parts of Africa. The vegetable oil extracted from melon seed is primarily used for domestic consumption although medicinal uses [1, 2] and industrial applications for biodiesel [3, 4], soap, detergents and margarine production, have been reported. In addition, the seed cake is a rich source of protein and minerals for livestock feeds and nutritional supplements [5]. Conversely, the extraction of vegetable oil from melon seeds produces enormous quantities of lignocellulosic waste commonly called *melon seed husk* (MSH). Currently, MSH is combusted in open air or dumped in refuse dumps and landfill sites thereby exacerbating the prevalence of greenhouse gases (GHG) and environmental burden of pollutant emissions on the environment [6, 7]. The utilization of MSH for water treatment, compost material, soil remediation, and weed suppressants is widely reported in literature [8-10].

Consequently, there is an urgent need to explore more efficient routes for the valorization of agricultural waste such as MSH for future clean bioenergy applications. The lignocellulosic nature of MSH indicates it is a potentially sustainable solid biofuel feedstock (SBF) for solid (biochar), liquid and gaseous biofuels production via torrefaction, pyrolysis, and gasification technologies [11]. However, the utilization of MSH for clean biofuels, green chemicals and other bioenergy applications has not been exploited comprehensively. This is largely due to the lack of comprehensive understanding on thermochemical fuel properties of lignocellulosic waste essential for design calculations and process optimization of thermochemical conversion processes. Thermogravimetric analysis (TGA) is one of the most widely utilized analytic techniques for investigating the thermochemical fuel properties and thermal decomposition behaviour of lignocellulosic waste. Data from TGA can be employed by scientists, engineers and policy makers to efficiently make decisions regarding feedstock materials, cost analysis, process equipment design and optimization of biomass thermal conversion systems.

Therefore, this study is aimed at investigating the thermal decomposition kinetics of MSH under pyrolysis conditions using thermogravimetric analysis (TGA). The study also presents an overview of the thermochemical fuel properties of MSH and its potential as a sustainable solid biofuel feedstock for future applications in biomass thermal conversion systems.

## Experimental

*Materials and Methods*

Melon seeds were purchased from a local market in Nigeria and dehulled to obtain the husks investigated in this study. The dry melon seed husks (MSH) were milled in rotary dry miller and sieved using a 250 μm sieve to obtain homogenous sized particles. The pulverised MSH was subsequently characterised via ultimate analysis, proximate analysis and bomb calorimetry to determine the elemental composition, fuel properties and calorific value respectively. The ultimate analysis was determined in the Vario EL Microcube™ CHNS analyser using ASTM D5291 method. Proximate analysis was investigated by ASTM standard techniques E871, E872 and D1102 for moisture, volatiles and ash content respectively while fixed carbon was determined by difference. The heating value of the fuel was determined using an IKA C2000 bomb calorimeter. The results for thermochemical properties of MSH are presented in Table 1.

**Table 1**: Proximate and Ultimate analysis of MSH

| Element | Symbol | Composition (wt %) |
|---|---|---|
| Carbon | C | 51.80 |
| Hydrogen | H | 6.67 |
| Nitrogen | N | 0.85 |
| Sulphur | S | 0.40 |
| Oxygen | O | 40.28 |
| Heating Value (HHV) | HHV MJ/kg | 21.78 |
| Moisture | M | 7.53 |
| Volatile Matter | VM | 79.52 |
| Ash | A | 1.50 |
| Fixed Carbon | FC | 18.99 |
| Volatiles-Fixed carbon ratio | VM/FC | 4.19 |

The thermal analysis of MSH was examined by thermogravimetric analysis in the Netzsch 209 F3 TG analyser. Approximately 8 mg of MSH was heated in an aluminum crucible from room temperature to 800 °C and heating rates of 5, 10, 20 °C per min under nitrogen atmosphere. The thermograms of MSH were analysed using the proprietary Proteus™ Netzsch thermal analysis software. The resulting TG-DTG data (not presented here) was applied to the *Kissinger,* and *Flynn-Wall-Ozawa (FWO)* models to investigate to determine the kinetic parameters for MSH under pyrolysis conditions for degrees of conversion, $\alpha = 0.15$ to $0.60$.

*Kinetic theory*

Biomass pyrolysis involves the thermal conversion of biomaterials in the absence of air or oxygen to produce liquid, gas and solid (biochar). The general equation of biomass pyrolysis is described by the generic equation [11];

$$C_nH_mO_p (Biomass) \xrightarrow{Heat} \sum_{liquid} C_xH_yO_z + \sum_{gas} C_aH_bO_c + H_2O + C\ (char) \quad (1)$$

The one step global model for the pyrolysis of biomass is assumed to occur according to Eq 1. Conversely, the rate of pyrolytic decomposition of biomass can be described by the expression;

$$\frac{d\alpha}{dt} = k(T)f(\alpha) \quad (2)$$

Where *a* - degree of conversion; *t*, time; *f(α)* the reaction model and *k(T)* is temperature dependent rate constant given by the Arrhenius relation expressed as;

$$k(T) = Ae^{-E/RT} \quad (3)$$

Where the terms; *E, T, R,* and *A* represent the activation energy in $kJ\ mol^{-1}$, absolute temperature in K, molar gas constant in $J\ mol^{-1}K^{-1}$ and frequency factor in $min^{-1}$.

Furthermore, the reaction model and its derivative form *f(α)* can be applied to describe the pyrolytic decomposition of biomass assuming a reaction order denoted as, *n*, based on the expression;

$$f(\alpha) = (1-\alpha)^n \quad (4)$$

Consequently, the weight loss (α) of the biomass materials during thermal decomposition is defined by the expression;

$$\alpha = \frac{m_i - m_a}{m_i - m_f} \quad (5)$$

Where the terms $m_i$, $m_a$, and $m_f$ represent the initial mass, actual mass at time, *t*, and final mass of the biomass material at the end of the process, respectively.

Hence, by substituting Eq.s 3 and 4 into Eq.2 an expression for the pyrolysis of biomass materials can be deduced;

$$\frac{d\alpha}{dt} = Ae^{-E/RT}(1-\alpha)^n \quad (6)$$

For the thermal decomposition of biomass materials at linear heating rate, β = dT/dt, under non-isothermal thermogravimetric analysis, Eq. 6 can be re-written as;

$$\frac{d\alpha}{(1-\alpha)^n} = \frac{A}{\beta}e^{-E/RT}dT \quad (7)$$

Consequently the kinetics of biomass pyrolytic decomposition can be analysed from the numerical solutions of integrating Eq.7, using the model free kinetics of *Flynn-Wall-Ozawa* (FWO) [12].

*Kissinger Method*

The Kissinger method can be utilised to determine the kinetic parameters by plotting the terms In $(\beta/T^2_m)$ against $1000/T_m$ given in this expression [13];

$$In\left(\frac{\beta}{T_m^2}\right) = In\left(\frac{AR}{E}\right) - \frac{E}{RT_m} \tag{8}$$

The kinetic parameters $E$ and $A$ can be determined from the slope $-E/R$ and intercept $In\ (AR/E)$ of the curve respectively.

*Flynn-Wall-Ozawa (FWO) Method*

The FWO method determines the kinetic parameters from the plots of In ($\beta$) against $1000/T$ presented in the expression [12, 14];

$$In(\beta) = In\left(\frac{A_\alpha E_\alpha}{Rg_\alpha}\right) - 5.331 - 1.052\frac{E_\alpha}{RT_{\alpha i}} \tag{9}$$

Consequently, the activation energy E and the frequency factor A can be deduced from the slope given by $-1.052E/R$ and intercept of the curve, respectively.

**Results and Discussion**

*Kinetic Analysis*

Figures 1 and 2 present plots for the kinetic analysis of MSH using the *Kissinger* and *Flynn-Wall-Ozawa* methods, respectively. By applying Eq. 8, the kinetic parameters $E$ and $A$ for MSH were calculated from the peak decomposition temperatures; 332.60 °C, 346.50 °C and 358.20 °C of MSH at the heating rates of 5, 10, 20 °C per min using the *Kissinger* method.

Consequently, the kinetic parameters $E$ and $A$ were deduced from the slope and intercept of the plot lines in Fig 1. The values for activation energy, $E = 161.26$ kJ/mol and frequency factor, $A = 2.08 \times 10^{10}$ min$^{-1}$ with the correlation value, $R^2 = 0.9958$. The Kissinger method has been applied in literature to investigate the thermokinetics of biomass [15].

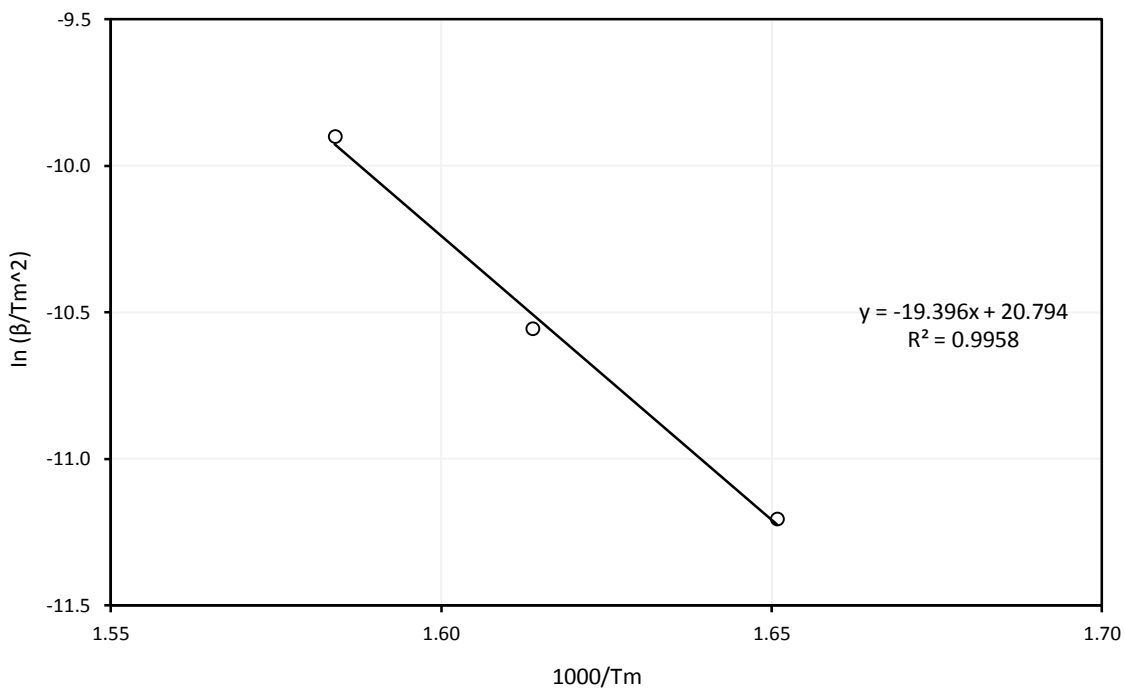

**Figure 1**: Kinetic plots for the *Kissinger* method.

Figure 2 represents the kinetic plots for MSH using the FWO method for degrees of conversion α = 0.05 to 0.60 at heating rates of 5, 10, and 20 °C per min from room temperature to 800 °C.

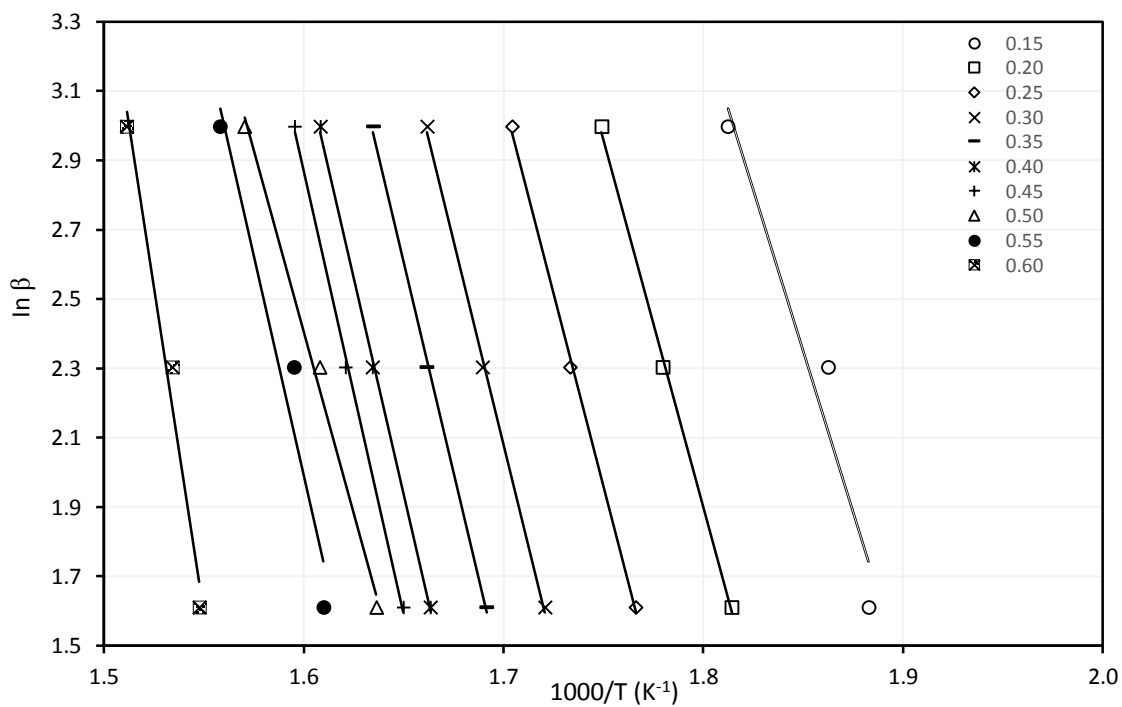

**Figure 2**: Kinetic plots for *Flynn-Wall-Ozawa* (FWO) method.

As can observed Fig. 2, the kinetic plots for MSH for the degrees of conversion = 0.5 to .60 using the FWO method are parallel to each other which suggests that the order and mechanism of thermal decomposition reactions of MSH are similar. It is generally agreed that the one step global model for biomass kinetic decomposition is assumed to be first order [16] as defined in Eq. 1 [11]. Table 2 presents the kinetic parameters for MSH calculated from the slope and intercept of the kinetic plots of FWO in Fig.2 obtained at high correlation values.

**Table 2**: Kinetic parameters for MSH from the FWO method.

| α | $R^2$ | E (kJ/mol) | A (min$^{-1}$) |
|---|---|---|---|
| 0.15 | 0.9422 | 146.81 | 1.56 x 10$^{17}$ |
| 0.20 | 0.9988 | 167.35 | 4.81 x 10$^{18}$ |
| 0.25 | 0.9989 | 176.42 | 1.39 x 10$^{19}$ |
| 0.30 | 0.9989 | 185.73 | 4.02 x 10$^{19}$ |
| 0.35 | 0.9989 | 192.06 | 8.14 x 10$^{19}$ |
| 0.40 | 0.9989 | 198.51 | 1.65 x 10$^{20}$ |
| 0.45 | 0.9980 | 201.77 | 2.35 x 10$^{20}$ |
| 0.50 | 0.9933 | 165.39 | 7.68 x 10$^{16}$ |
| 0.55 | 0.9416 | 199.58 | 6.22 x 10$^{19}$ |
| 0.60 | 0.9773 | 296.00 | 2.86 x 10$^{27}$ |
| **Average** | **0.9847** | **192.96** | **2.86** x 10$^{26}$ |

The values of E for the fuel observably increased from α = 0.15 (E = 146.81 kJ/mol) to α = 0.45 (E = 201.77 kJ/mol) decreased at 0.5 before increasing to 296 kJ/mol which marks the highest activation observed during the thermal analysis of MSH. This indicates that the decomposition of MSH is slowest at α = 0.60 while the process is fast at α = 0.15. Overall the average value for activation energy, E is 192.96 kJ/mol, frequency factor, A is 2.86 x 10$^{26}$ min$^{-1}$, at correlation values of 0.9847.

In comparison, the average kinetic values of MSH using FWO are clearly higher than values from the *Kissinger* method E = 161.26 kJ/mol and frequency factor, A = 2.08 x 10$^{10}$ min$^{-1}$ with the correlation value, $R^2$ = 0.9958. This may be due to the fact that the *Kissinger* method only accounts for the maximum peak decomposition temperature ($T_{max}$) of the fuel compared to extensive temperature range (based on degree of conversions) of the isoconversional FWO method. Overall, the parametric kinetic values of MSH fall are in good agreement with the range of values reported for biomass in literature [17].

## Conclusion

The study was aimed at investigating the thermochemical fuel and kinetic decomposition characteristics of melon seed husks (MSH) under inert (pyrolysis) conditions. The elemental composition, proximate analyses, calorific value and thermal decomposition kinetics of MSH was examined using the *Kissinger* and *Flynn-Wall-Ozawa* methods. The results indicated that the average parametric kinetic values of MSH using FWO method were higher than the *Kissinger* method. This can be ascribed to the isoconversional nature of the FWO method over an extensive temperature range (based on degree of conversions) compared to the maximum peak decomposition temperature ($T_{max}$) of *Kissinger* method. The parametric kinetic values were observed to be in good agreement with literature values for other biomass. The findings of the study also indicate that MSH possesses important characteristics of a potential solid biofuel (SBF) for utilization in future thermochemical applications for clean bioenergy generation.